\documentclass[runningheads]{rnl}
\usepackage[latin1]{inputenc}
\usepackage[francais]{babel}

\usepackage{graphicx,epsf,subfigure}  

\begin{document}
 
\title*{Chaos généré par une non linéarité 2D et une dynamique à
  retard}
 
\titretab{Générateur de chaos opto--électronique à double retard
  \protect\newline pour les télécommunications optiques sécurisées}
 
\titrecourt{Non linéarité 2D et dynamique à retard}

 
\author{ Mourad Nourine \and Michael Peil \and Laurent Larger }

\index{Nourine Mourad} \index{Larger Laurent} \index{Peil Michael}

\auteurcourt{M. Nourine, M. Peil \& L. Larger}
 
\adresse{Département d'Optique, Institut FEMTO--ST, UMR CNRS 6174,
  Université de Franche--Comté, 25030 Besançon Cedex, France.}
\email{mourad.nourine@univ-fcomte.fr}
\maketitle



\begin{resume}
  Nous présentons dans cet article une nouvelle architecture optoélectronique pour la génération de chaos en intensité. Le principe s'appuie sur une dynamique électro-optique non linéaire à  retard, dont la non linéarité est construite grâce à un interféromètre à 4 onde réalisé en optique intégrée, et disposant de  2 électrodes de modulation indépendantes. Le montage permet de  disposer d'une part, d'une dynamique ultra-rapide jusqu'à des
  fréquences de plusieurs GHz, et d'autre part, de générer un chaos de grande dimension destiné au cryptage physique de données optiques.  Au travers d'une étude numérique et expérimentale, nous avons  cherché à analyser certains des nombreux comportements dynamiques  que peut présenter cet oscillateur, en fonction de divers paramètres physiques du montage~: régimes de point fixes stables, périodiques, et chaotiques. La mise en {\oe}uvre du montage expérimental a permis de valider le modèle théorique adopté pour les simulations.
\end{resume}


\begin{resumanglais}
  We present a new optoelectronic architecture intended for chaotic optical intensity generation. The principle relies on an electro-optic non-linear delay dynamics, which non linearity is performed by a 4-waves integrated optics interferometer involving 2  independent electro-optic modulation inputs. The setup allows both to have an ultra-fast dynamics up to several GHz frequencies, and potentially high dimensional chaos intended for encryption of optical data at the physical layer. We have built a mathematical  model of the system and analyzed a number of its possible solutions:
  stable steady states, periodic and chaotic regimes. The experimental observations allowed to validate the dynamical model, through good
  qualitative agreements with the numerical simulations.
\end{resumanglais}


\section{Introduction}

Les communications sécurisées par chaos ont permis le développement de nombreux systèmes dynamiques optiques produisant des comportements
chaotiques complexes \cite{Goedgebuer}.  La plupart de ces réalisations utilisent une classe particulière de comportements chaotiques, celle des dynamiques non linéaires à retard \cite{Ikeda}. Ces dynamiques temporelles possèdent la particularité étrange d'évoluer dans un espace des phase de dimension infinie, dans lequel des comportements chaotiques de grande complexité peuvent être observés \cite{Farmer}. L'oscillateur électro-optique que nous allons présenter fait partie de cette catégorie, et le chaos est observée sur la variable intensité optique.

Tout d'abord, le dispositif expérimental sera décrit, ce qui permettra d'en établir un modèle théorique sous forme d'un système de deux équations différentielles couplées du second ordre, non linéaires, et à retards multiples. Ce modèle sera par la suite utilisé pour effectuer des simulations numériques. Ensuite, nous présenterons et comparerons les résultats obtenus par des mesures expérimentales et ceux obtenus numériquement.  Enfin, nous conclurons en décrivant le système cryptographique complet, qui doit à terme utiliser ce générateur de chaos comme dispositif émetteur d'un côté, et comme récepteur~/ décodeur de l'autre côté.
  
\section{Générateur de chaos à modulateur QPSK}

\subsection{Description et principe de fonctionnement}

\begin{figure}[htbp]
  \begin{center}
    \epsfysize=6cm \leavevmode \epsfbox{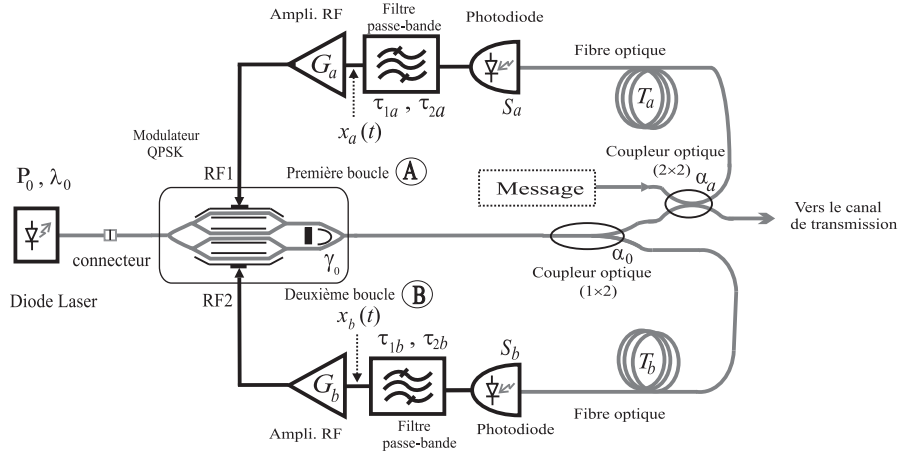}
    \caption{Schéma du dispositif expérimental d'un générateur de
      chaos à modulateur QPSK.}
    \label{Nourinefig1}
  \end{center}
\end{figure}

Le dispositif expérimental de l'oscillateur chaotique est illustré sur la figure \ref{Nourinefig1}. Le générateur est formé par 2 boucles de contre-réaction reliées à un modulateur QPSK (Quadrature Phase Shift Keying). Ce modulateur électro-optique est particulier par son architecture. Il appartient à la famille des modulateurs Mach-Zehnder (MZ) à 4 bras, et il est intégré sur Niobate de Lithium (LiNbO$_3$). Cet élément clé est un composant commercial récent, originellement destiné à des nouveaux formats de modulation numérique pour les télécommunications optiques \cite{Noe}. Il permet pratiquement dans notre cas la réalisation d'une fonction non linéaire bidimensionnelle.

\vspace{-0.5cm}

\noindent Le reste du générateur de chaos est plus classique.  Il est constitué (de gauche à droite) par les composants suivants~:

\begin{itemize}
\item une diode laser monomode reliée à une fibre optique à maintien   de polarisation~; sa longueur d'onde d'émission est celle   habituellement utilisée dans les télécommunications optiques ($\lambda_0=1,55\;\mu$m).
\item le modulateur électro-optique QPSK déjà éviqué. Il a la  particularité d'avoir $2$ entrées RF indépendantes et $3$ tensions  de polarisation indépendantes. Ce composant représente le c\oe ur du système, sa fonction de transfert de modulation réalisant la fonction non linéaire du générateur de chaos.
\item le système présente aussi l'originalité d'avoir $2$ boucles de contre-réactions électro-optiques non linéaires, impliquant chacune un élément retardant, une photodiode de conversion optique~/ électrique, un filtre électronique large bande responsable des   termes différentiels de la loi, et un élément amplificateur.
  \begin{description}
  \item [1. La première boucle] est réalisée par la mise en série des éléments suivants~:
    \begin{itemize}
    \item un coupleur optique (2 entrées~/ 2 sorties) permettant d'insérer le message utile par l'une de ces entrées, et d'envoyer le mélange (chaos + message) d'une part vers la première contre-réaction, et d'autre part vers la canal de transmission.
    \item une fibre optique réalisant un retard temporel pur, et constituant un paramètre physique \og clé \fg $\;$ de la première boucle de contre-réaction (La valeur très précise de ce  retard est nécessaire pour l'opération de synchronisation~/ décodage).
    \item une photodiode permettant de convertir la puissance optique en un signal électrique.
    \item un filtre passe bande limitant la bande passante du signal chaotique électrique.
    \item un amplificateur électronique réglable pour ajuster le gain de boucle de la contre-réaction.
    \end{itemize}
  \end{description}
  \begin{description}
  \item [2. La deuxième boucle] de rétroaction est réalisée par des éléments semblables à ceux utilisés dans la première boucle, à l'exception du coupleur optique à (2 entrées~/ 2 sorties) qui ne fait plus partie de cette seconde boucle de contre-réaction.
  \end{description}
\end{itemize}

Le processus dynamique réalisé par cet oscillateur en double boucle fermée peut être décrit de la manière suivante~: une diode laser fibrée alimente optiquement le modulateur QPSK. Le faisceau lumineux traversant le modulateur QPSK subit à la sortie une variation d'intensité non linéaire par rapport à chacune des deux tensions appliquées aux électrodes de commandes RF1 et RF2 du modulateur. Le coupleur (1$\;$entrée/2 sorties) divise ensuite cette intensité optique en deux quantités égales. Les signaux issus des sorties de ce coupleur sont ensuite retardés différemment dans chaque boucle par une
certaine longueur de fibre optique (quelques centimètre à quelques mètres), puis ils sont détectés par des photodiodes. Les signaux électriques obtenus sont filtrés, puis amplifiés pour y être réinjectés sur les électrodes de modulation RF du QPSK. La présence du coupleur optique (2 entrées~/ 2 sorties) dans la première boucle ne sert qu'à mélanger le message informatif avec la modulation chaotique de l'intensité.
  
\subsection{Modélisation de la fonction non linéaire }

\begin{figure}[htbp]
  \begin{center}
    \begin{tabular}{ccc}
      \setlength{\epsfysize}{3.5cm}
      \subfigure[\scriptsize Schéma élémentaire \hspace{1cm}]{\epsfbox{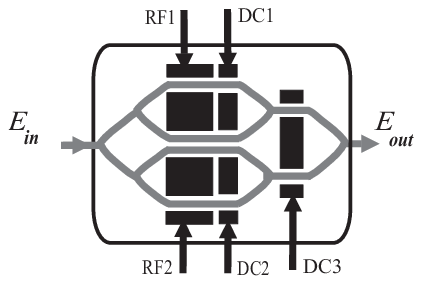}} &
      \setlength{\epsfysize}{3.5cm}
      \subfigure[\scriptsize Modélisation \hspace{1cm}]{\epsfbox{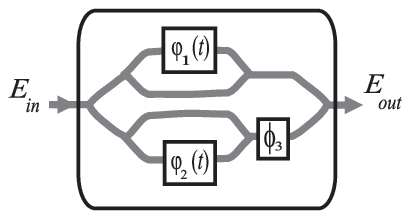}} &
      \setlength{\epsfysize}{3.5cm}
      \subfigure[\scriptsize Allure de la fonction de transfert]{\epsfbox{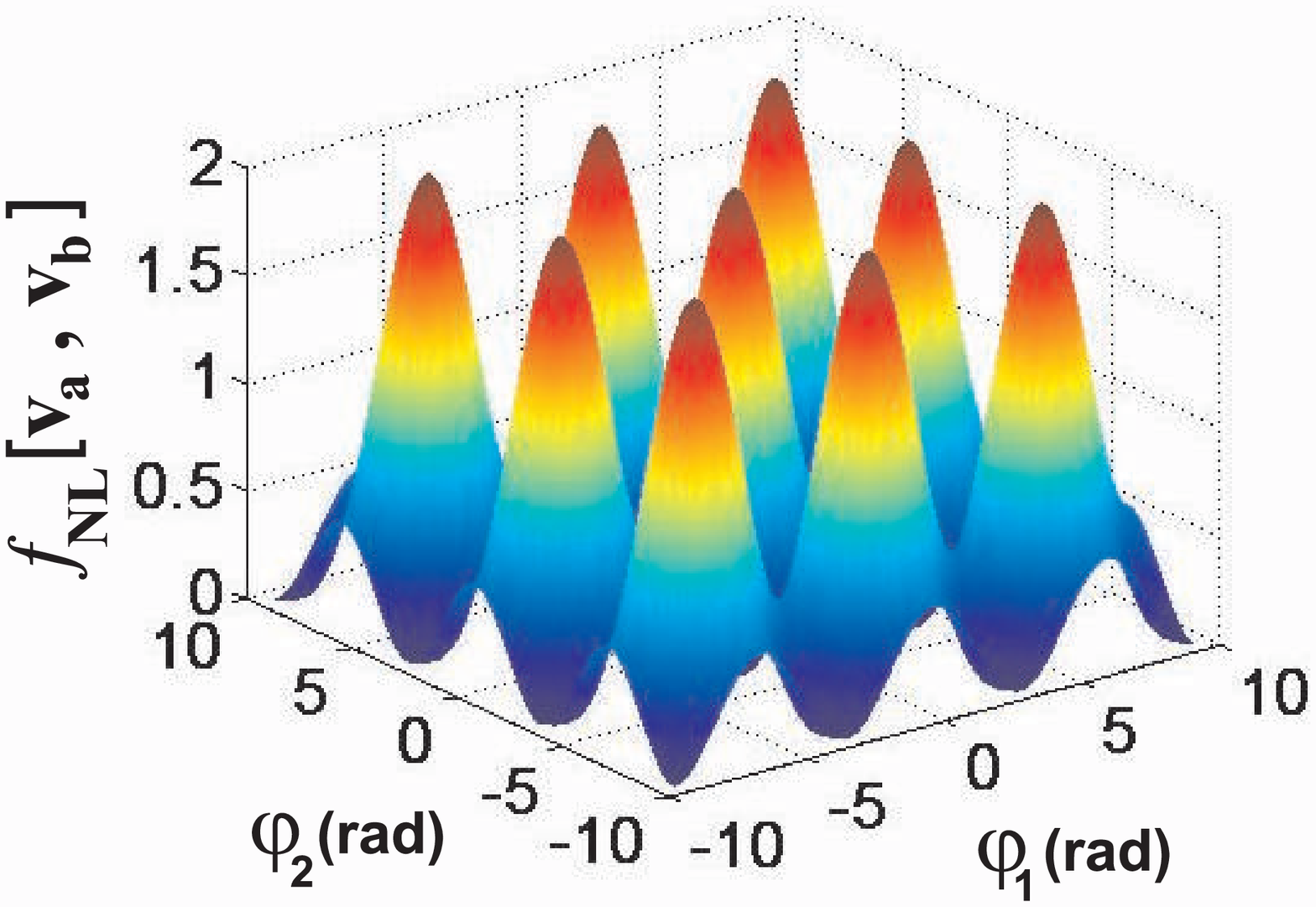}} \\[-0.2cm]
    \end{tabular}\\[0.1cm]
    \caption{La fonction non linéaire réalisée par le modulateur QPSK.}
    \label{Nourinefig2}
  \end{center}
\end{figure}

La figure \ref{Nourinefig2}.a schématise la configuration du modulateur QPSK utilisé pour la réalisation de la fonction non linéaire bidimensionnelle, notée par: $f_{N\!L}[v_a, v_b]$. Cette dernière est représentée par la fonction de transfert du modulateur. L'expression de l'intensité optique, qui représente la variable de sortie du modulateur QPSK, est fonction des tensions appliquées sur ces électrodes RF ($v_a$ sur RF1 et $v_b$ sur RF2) ainsi que de trois tensions continues (DC1, DC2 et DC3) qui permettent d'ajuster des points de repos de la condition d'interférence à onde multiple.

\noindent L'expression analytique de la fonction de transfert du modulateur QPSK peut s'établir de la manière suivante:

\begin{itemize}
\item Sur l'électrode DC$_{m}$ ($m=1, 2,3$) est appliquée une tension   continue $V_{DC_m}$. Un déphasage $\phi_m$ est introduit sur l'onde
  optique qui se propage le long de la branche soumise à la tension  $V_{DC_m}$ (comme illustré sur la figure \ref{Nourinefig2}.b).   L'expression de ce déphasage est donnée par~:
  \begin{equation}
    \label{Nourineeq1}   
    \phi_m = \pi\cdot\frac{V_{DC_{m}}}{V_{\pi DC_{m}}}
  \end{equation}	
  où $V_{\pi DC_m}$ est la tension demi-onde, qui permet de réaliser un déphasage de $\pi$.
	
\item Sur les électrodes RF$_{1,2}$ sont appliquées respectivement 2 tensions variables $v_{a,b}(t)$.  Ces dernières sont des tensions de   modulation, et elles introduisent des déphasages variables $\varphi_{1,2}(t)$.  Les expressions de ces déphasages sont données par~:
  \begin{equation}
    \label{Nourineeq2}   
    \varphi_{1,2}(t) = \pi\cdot\frac{v_{a,b}(t)}{V_{\pi RF_{1,2}}} +  \phi_{1,2} 
  \end{equation}
  où $V_{\pi RF_{1,2}}$ sont des tensions demi-ondes, qui permettent de réaliser en régimes dynamiques un déphasage de $\pi$   respectivement dans les interféromètres de Mach-Zehnder MZ$_1$ et MZ$_2$.
\end{itemize}

\noindent L'expression de l'intensité optique $I_{out}(t)$ à la sortie du modulateur QPSK est obtenue par le calcul de la moyenne (sur un temps défini par le temps de réponse électronique) du module au carré du champ électrique résultant $E_{out}(t)$~:
\begin{equation}
  \label{Nourineeq3}   
  I_{out}(t)=\langle|E_{out}(t)|^{2}\rangle= f_{N\!L}[v_{a},v_{b}](t)
\end{equation}

En imposant un champ en entrée du modulateur sous la forme $E_{in}(t)=\sqrt{P_0}.\exp(i\omega_0 t)$ ($P_0$ est la puissance optique à l'entrée du QPSK, $\omega_0=2\pi.c/\lambda_0 $ est la fréquence angulaire de la source laser, et $c$ est la vitesse de la lumière), le champ électrique $E_{out}(t)$ en sortie du modulateur QPSK est donné par~:

\begin{equation}
  \label{Nourineeq4}   
  E_{out}(t)=\frac{\sqrt{P_0}}{2}\cdot\biggl
  [1+\exp{\bigr(i\varphi_{1}(t)\bigr)}+ \Bigl[1+
  \exp{\bigr(i\varphi_{2}(t)\bigr)}\Bigr]\cdot
  \exp{(i\phi_3)}\biggr]\cdot\exp{(iw_{0}t)}
\end{equation}

\noindent Il vient finalement pour la fonction de modulation non linéaire bidimensionnelle~:

\begin{equation}
  \label{Nourineeq5}  
  f_{N\!L}[v_{a},v_{b}](t)=
  \frac{P_0}{2}\biggr\{\cos(\psi_3)\cdot\Bigl[\cos(\psi_3)+
  2\cdot\cos\bigl(\psi_1+\psi_2\bigl) 
  \cdot\cos\bigl(\psi_2+\psi_3 -\psi_1\bigl) \Bigl]
  +\cos^{2}\bigl(\psi_2+\psi_3-\psi_1\bigl)\biggr\}
\end{equation}
\hspace{2cm}avec: \vspace{-0.5cm}
\begin{eqnarray}
  \psi_1=\frac{\varphi_{1}(t)}{2};\hspace{1cm}\psi_2
  =\frac{\varphi_{2}(t)}{2}; \hspace{1cm}\psi_3=\frac{\phi_{3}}{2};
  \nonumber
\end{eqnarray} 

\noindent Une condition suffisante sur la non linéarité pour obtenir une dynamique chaotique est de présenter un extrémum dans l'intervalle de variation des variables d'entrées. Cette condition est suffisamment vérifiée par $f_{N\!L}[v_{a},v_{b}]$, donnée par la relation (\ref{Nourineeq5}), comme le montre un exemple de son allure sur la figure$\;$\ref{Nourinefig2}.c (les paramètres utilisés pour tracer cette figure sont donnés au tableau \ref{NourineTab1}).

\section{Modélisation du système }

Le processus différentiel se modélise à partir de la fonction de transfert du filtre passe-bande de la branche électronique de l'oscillateur.  Comme l'oscillateur est formé par 2 boucles de contre-réactions contenant 2 filtres passe-bande différents, nous adoptons pour désigner les paramètres de chaque boucle les notations suivantes~: tous les paramètres de la boucle$\;$(A) seront indexés par la lettre \og a\fg, et tous les paramètres de la boucle (B) seront indexés par la lettre \og b\fg. Ainsi, la dynamique globale est modélisée par 2 équations différentielles du second ordre à retard
(système d'EDR), dont leurs expressions sont données par~:

\begin{equation}
  \label{Nourineeq7}
  x_i(t)+[\tau_{1i}+\tau_{2i}]\frac{dx_i}{dt}(t)+
  \tau_{1i}\cdot\tau_{2i}\frac{d^{2}x_i}{dt^{2}}(t)  = \displaystyle
  \beta_{i}\cdot \frac{d}{dt}\Bigl[f_{N\!L}[x_{a},x_{b}](t-T_i)\Bigl]
\end{equation}\medskip

\noindent où ($i=a,b$) selon la boucle concernée, et  $x_i(t)=v_i(t)/(2\cdot V_{\pi RFi})$ représentent les variables normalisées où $v_i(t)$. $\tau_{1i}$ et $\tau_{2i}$ sont les constantes de temps caractéristiques du profil des filtres électroniques, elles sont reliées directement aux fréquences de coupures $f_{c1i}$ et $f_{c2i}$ haute et basse du filtre. $\beta_{i}=(\pi\cdot P_0\cdot \gamma_0\cdot G_i\cdot S_i\cdot
\alpha_0\cdot \alpha_i)/(2\cdot V_{\pi RFi})$ est le gain global normalisé de la boucle de rétroaction, avec~: $\gamma_0$ le coefficient des pertes optiques du modulateur QPSK; $G_i$ le gain de l'amplificateur; $S_i$ la sensibilité du photodétecteur; $\alpha_0$ le coefficient de couplage du couleur optique ($1\times2$).  $\alpha_i$ est le coefficient de couplage du coupleur optique ($2\times2$) (qui n'apparaît pas pour la boucle$\:$(B)).

\section{Résultats numériques et expérimentaux}

Les simulations ont été effectuées en intégrant le modèle donné par (\ref{Nourineeq7}) par la méthode du prédicteur-correcteur. Un outil important dans la compréhension de la route vers le chaos est le diagramme de bifurcation, qui permet de voir rapidement l'ensemble des régimes dynamiques obtenus pour différentes valeurs du paramètre de bifurcation. Ce dernier correspond dans notre cas au gain normalisé $\beta_a$ de la contre-réaction de la boucle (A).

\begin{figure}[ht]
  \begin{center}
    \begin{tabular}{ccc}
      \setlength{\epsfysize}{3.55cm}
      \subfigure[expérimental]{\epsfbox{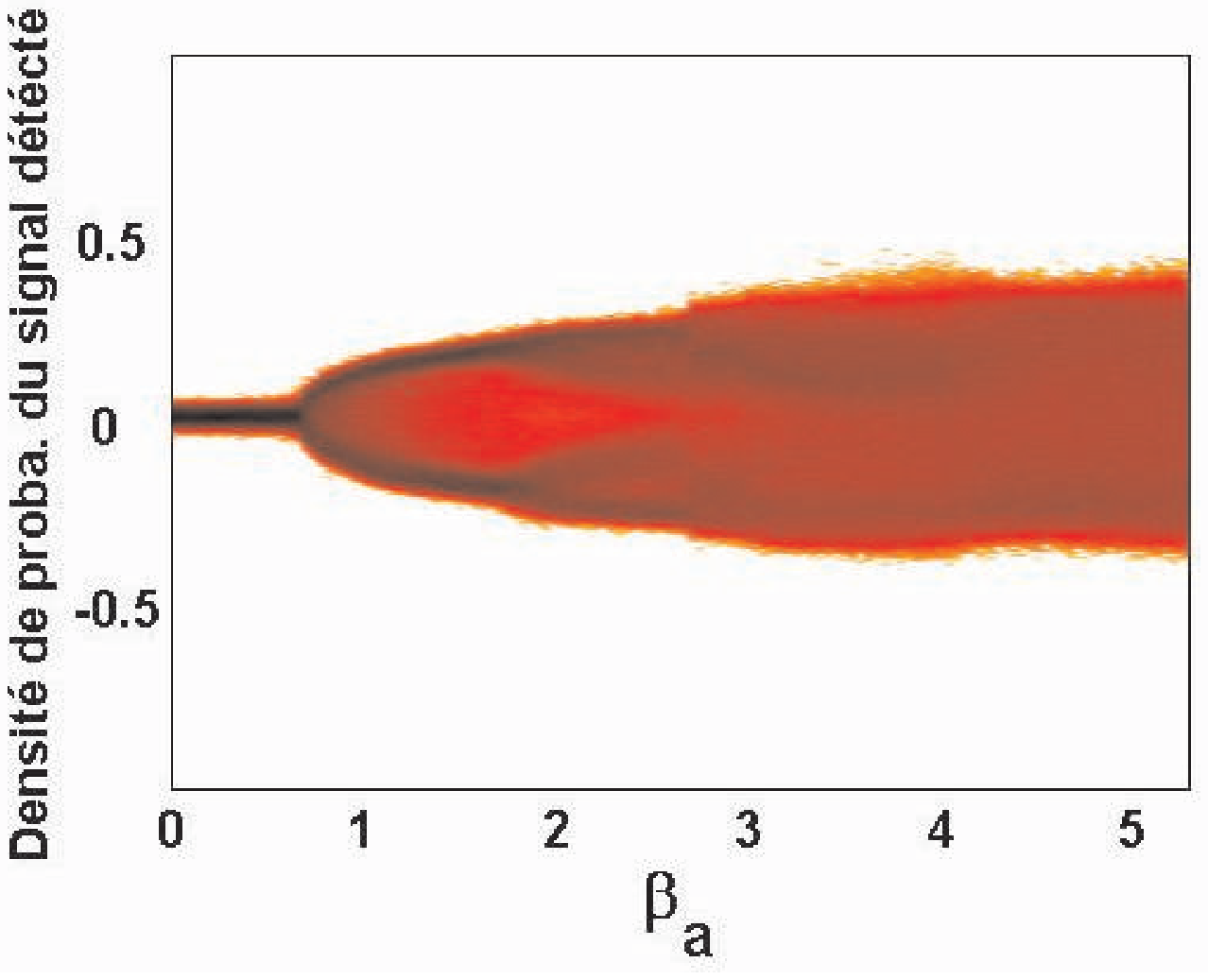}} &
      \setlength{\epsfysize}{3.55cm}
      \subfigure[simulation]{\epsfbox{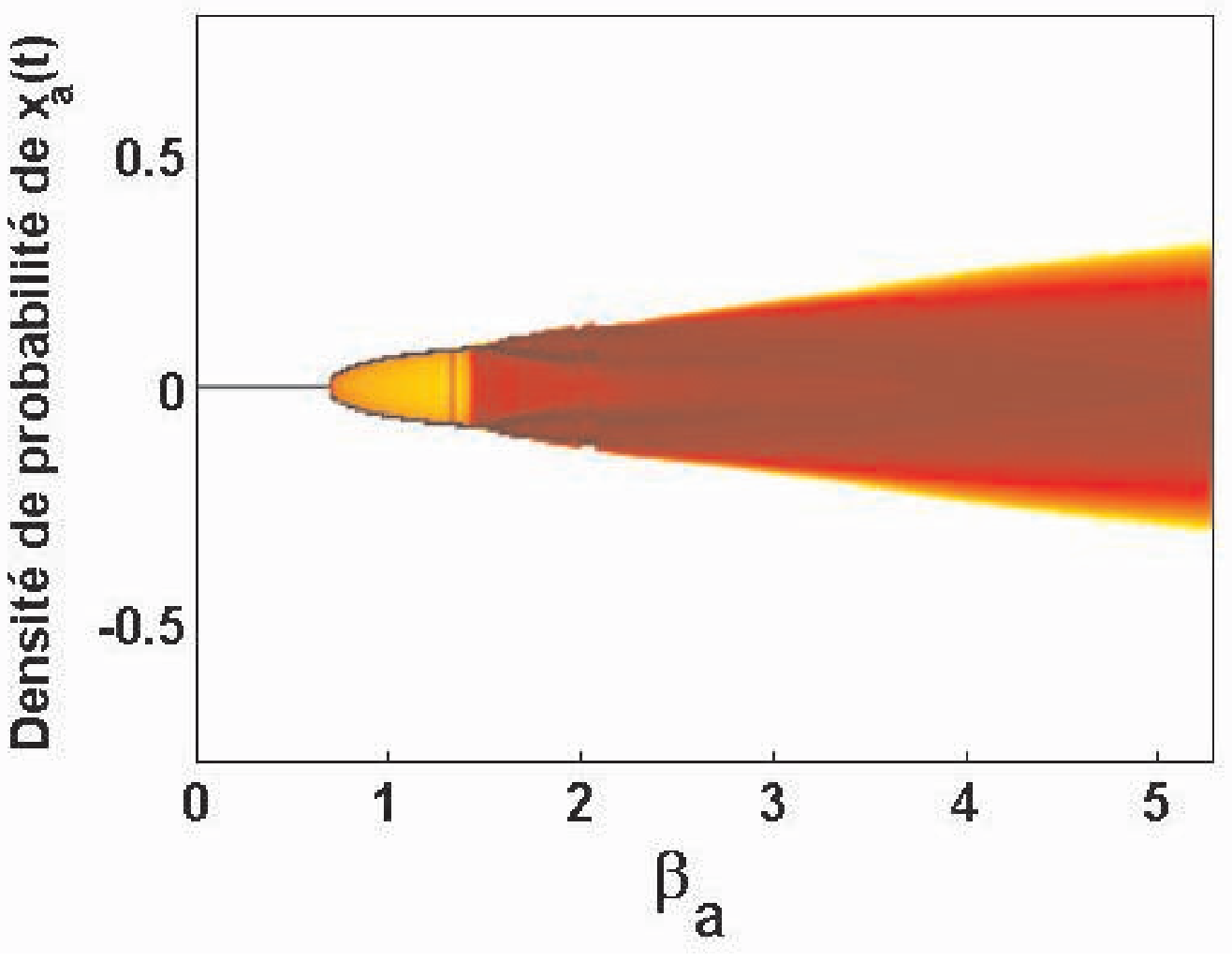}} &
      \setlength{\epsfysize}{3.55cm}
      \subfigure[entropie]{\epsfbox{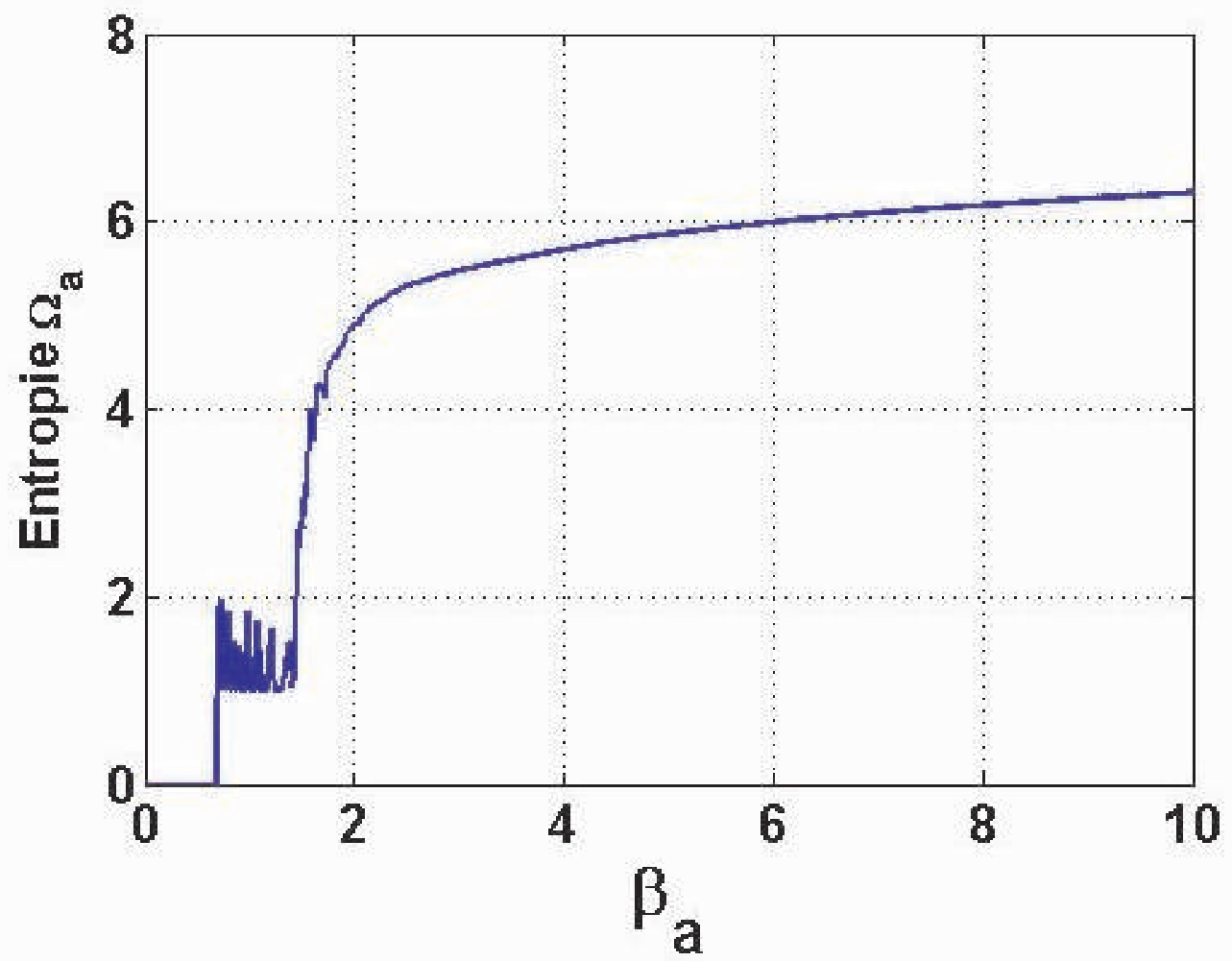}} \\[0.2cm]
    \end{tabular}\\[0.1cm]
    \caption{Diagrammes de bifurcation et entropique du système en une
      seule boucle de rétroaction ($\beta_b=0$).}
    \label{Nourinefig3}
  \end{center}

  \begin{center}
    \begin{tabular}{ccc}
      \setlength{\epsfysize}{3.5cm}
      \subfigure[expérimental]{\epsfbox{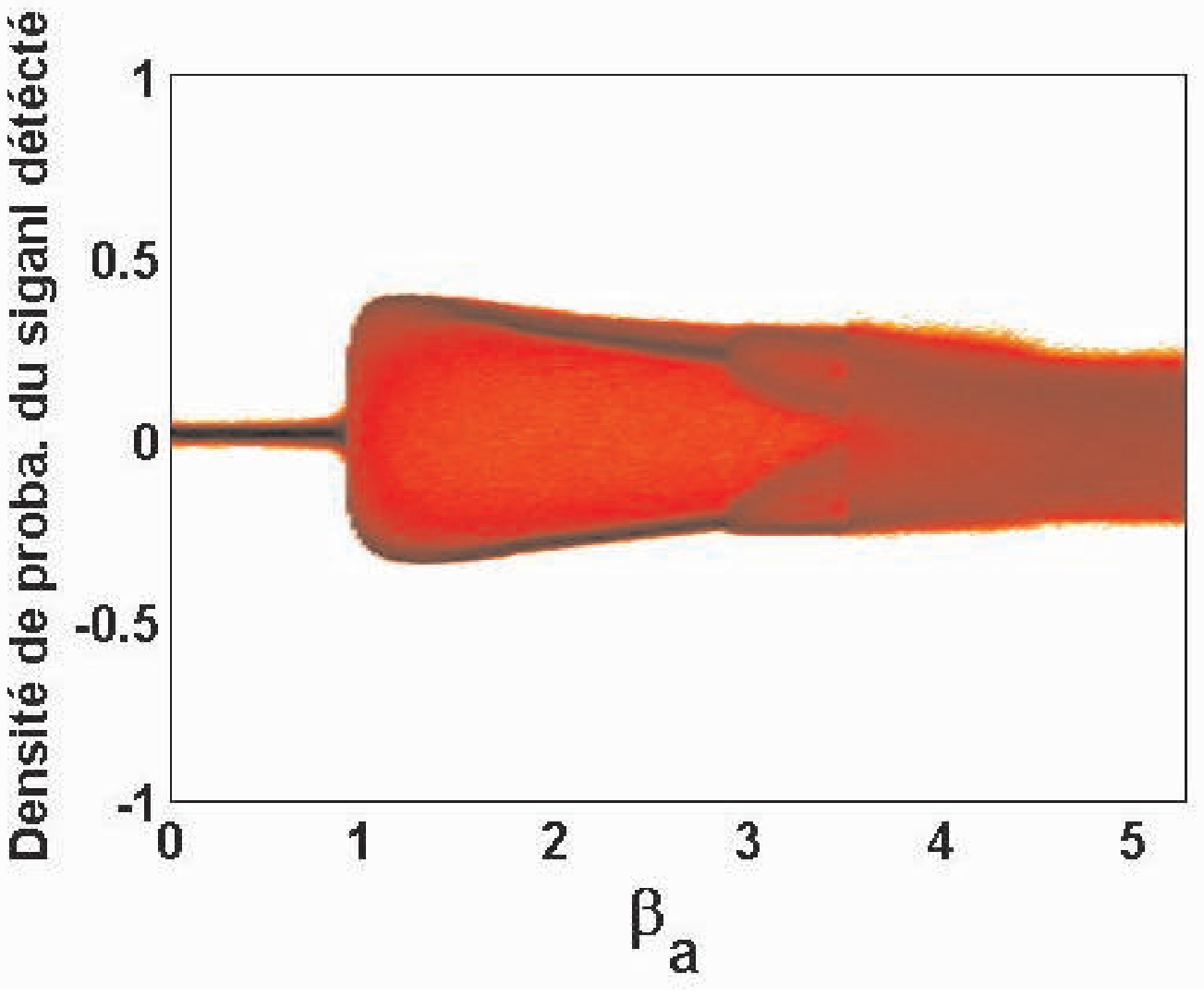}} &
      \setlength{\epsfysize}{3.4cm}
      \subfigure[simulation]{\epsfbox{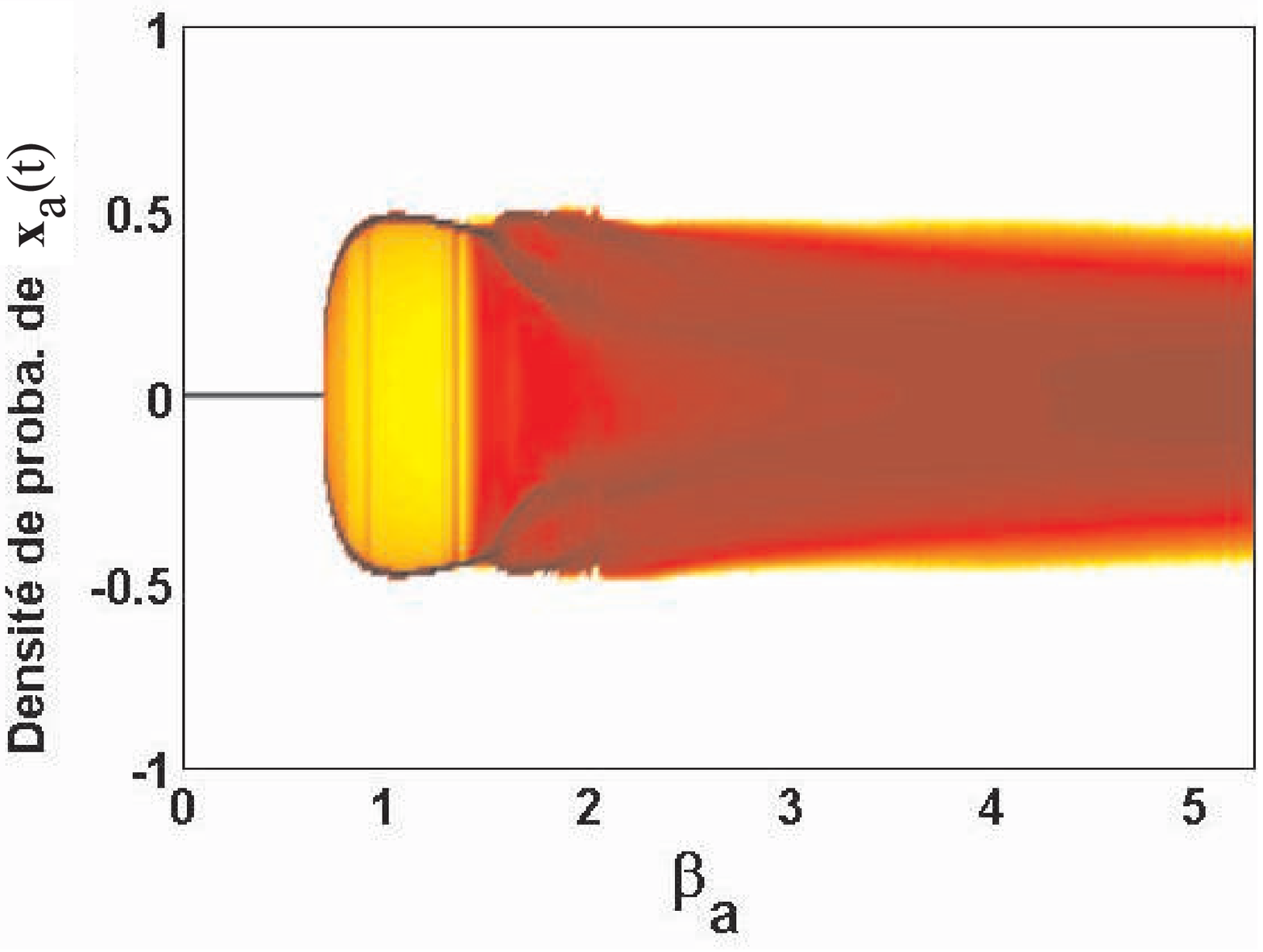}} &
      \setlength{\epsfysize}{3.6cm}
      \subfigure[entropie]{\epsfbox{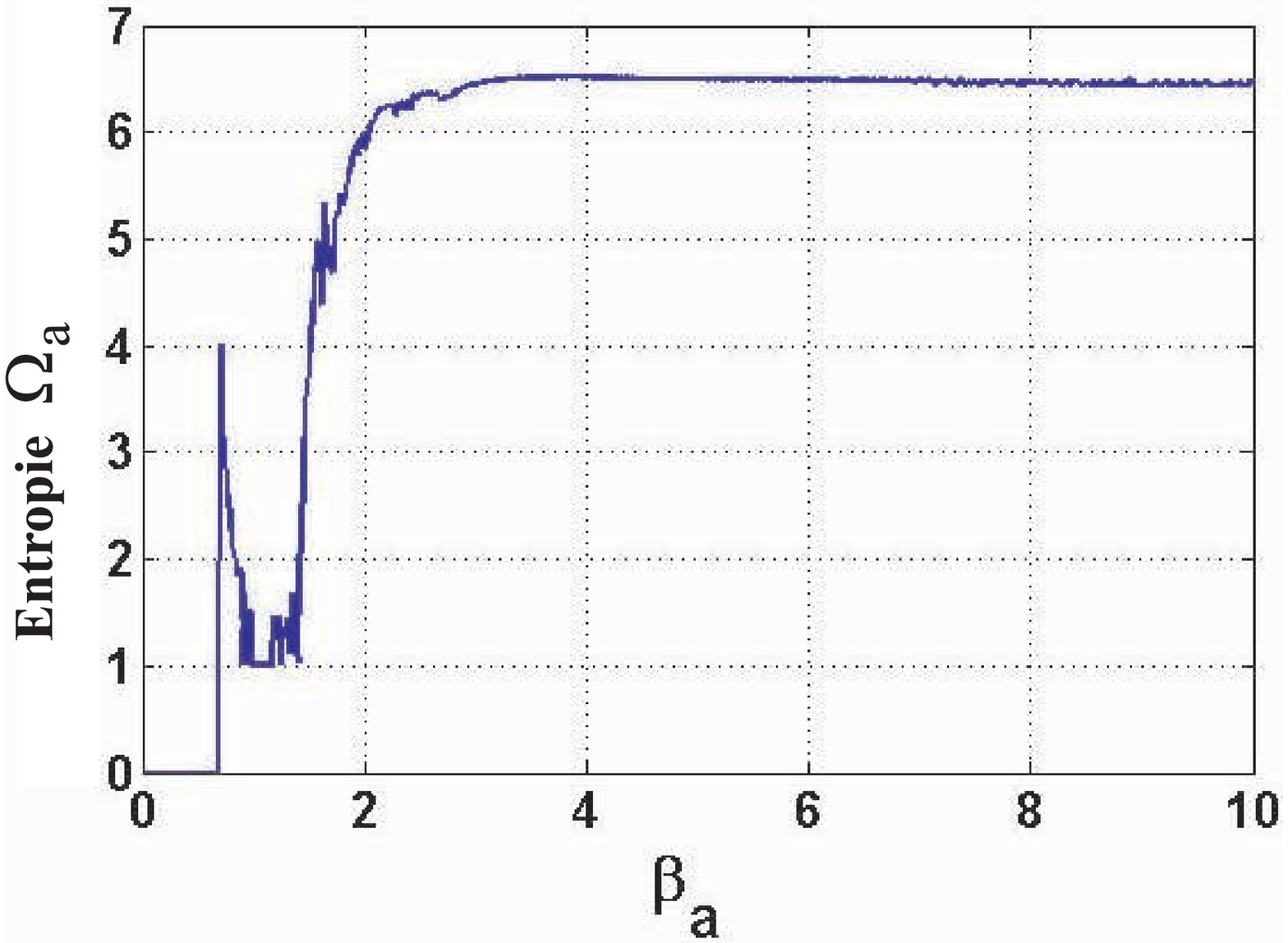}} \\[-0.2cm]
    \end{tabular}\\[0.1cm]
    \caption{Diagrammes de bifurcation et entropique du système en
      doubles boucles de rétroaction ($\beta_b=0,1$). }
    \label{Nourinefig4}
  \end{center}
\end{figure}

Sur les figures \ref{Nourinefig3}.a et \ref{Nourinefig3}.b, nous avons représenté les diagrammes de bifurcation obtenus expérimentalement et par simulation numérique. Afin de faciliter la comparaison expérience~/ simulation, le mode de représentation correspond à l'évolution, en fonction d'un gain de boucle, de la densité de probabilité de la variable normalisée $x_a$ (résultat du filtrage passe-bande de $f_{N\!L}[x_{a},x_{b}]$, dans la boucle (A)).\\ Des premiers diagrammes ont été effectués  dans le cas d'une seule boucle de rétroaction, cas qui peut servir de référence puisqu'il est déjà étudié dans la littérature \cite{Larger}. Le système en une seule boucle signifie que physiquement, une boucle est laissée ouverte, et  théoriquement, il se traduit par un gain nul ($\beta_b=0$). On remarque sur ces diagrammes l'évolution typique de la dynamique du système jusqu'au chaos selon une route du type cascade par dédoublements. Les dynamiques chaotiques obtenues se traduisent par une entropie élevée, et une densité de probabilité diffuse, et de profil quasi-Gaussien.

\begin{table}[ht]
  \begin{center}
    \begin{tabular}{|c c c||c c c|} \hline
      \multicolumn{3}{|c||}{\textbf{Boucle (A)}} &
      \multicolumn{3}{c|}{\textbf{Boucle (B)}} \\ \hline symbole &
      \hspace{0.5cm} valeur\hspace{0.5cm} & unité & symbole &
      \hspace{0.4cm} valeur \hspace{0.4cm} & unité \\ \hline
      $T_a$      &  61       & ns     &  $T_b$      & 60        & ns             \\
      $f_{c1a}$  &  13       & GHz    &  $f_{c1b}$  & 13        & GHz            \\
      $f_{c2a}$  &  50       & kHz    &  $f_{c2b}$  & 30        & kHz            \\[0.1cm]
      $\tau_{1a}=\frac{1}{2.\pi. f_{c1a}}$       & 12,2         & ps       & $\tau_{1b}=\frac{1}{2.\pi. f_{c1b}}$ &  12,2   & ps     \\[0.2cm]
      $\tau_{2a}=\frac{1}{2.\pi. f_{c2a}}$ & 3,18 & $\mu$s &
      $\tau_{2b}=\frac{1}{2.\pi. f_{c2b}}$ & 5,30 & $\mu$s \\ [0.2cm]
      \hline\hline \multicolumn{6}{|c|}{\small\textbf{Paramètres de la
          non linéarité}} \\ \hline
      $V_{\pi RF1}$      &  5,84       & V     &  $V_{\pi RF2}$    & 6,08      & V            \\
      $V_{\pi DC1}$      &  7,40       & V     &  $\phi_{1}$       & 2,9       & rad          \\
      $V_{\pi DC2}$      &  7,14       & V     &  $\phi_{2}$       & 1,3       & rad          \\
      $V_{\pi DC3}$ & 14,24 & V & $\phi_{3}$ & -0,1 & rad \\ \hline
    \end{tabular}
  \end{center}
  \caption{Paramètres expérimentaux du  générateur de chaos à modulateur QPSK.}{\label{NourineTab1}}
\end{table}

Les diagrammes de bifurcation du système en double boucle sont donnés sur les figures \ref{Nourinefig4}.a et \ref{Nourinefig4}.b. On remarque sur le diagramme de bifurcation expérimental (figure \ref{Nourinefig4}.a) les différents régimes dynamiques (régimes points fixes stables, régimes périodiques et chaotiques). Le diagramme obtenu par simulation reproduit le résultat expérimental avec une très bonne ressemblance (figure \ref{Nourinefig4}.b).

\section{Conclusion}

Dans cet article, nous avons présenté une nouvelle architecture de générateur de chaos en intensité, basé sur une dynamique non linéaire à retard. Le dispositif expérimental est réalisé par des composants électro-optiques, dont le c\oe ur du système est un modulateur QPSK. Ce modulateur permet de relier $2$ boucles de rétroactions, et d'introduire une non linéarité en 2 dimension. Cette architecture du générateur nous permet de disposer d'un nombre important de paramètres physiques (clés de codages) pour réaliser des communications sécurisées au niveau physique à haut débits.
   
Nous avons modélisé le système par un système d'équations différentielles du second ordre à retard. La comparaison des résultats de l'intégration de ce système par la méthode prédicteur-correcteur aux résultats expérimentaux, nous a permis de valider le modèle théorique proposé. Dans les simulations, nous avons pris soin de choisir un pas d'échantillonnage constant ($h=1$ ps) assez petit devant la plus petite constante de temps du système ($\tau_{1a}=\tau_{1b}=12,2$ ps).  Nous avons choisi aussi pour la réalisation du générateur, des retards temporels très grand devant la plus petite
constante de temps afin de garantir la génération d'un hyperchaos ($T_b/\tau_{1b}\approx 5000 \gg 1$), et d'optimiser par conséquent la confidentialité des transmissions.

Le travail va désormais se poursuivre sur la réalisation du système global de cryptographie par chaos, dont les premiers résultats numériques sont déjà obtenus en modulant chaotiquement une information binaire à plus de 3 Gbit/s. La restitution du message au niveau du récepteur est basée sur le principe de la synchronisation de chaos, initialement introduit par Pecora et Carroll \cite{Pecora}.


\vfill

\begin{thebibliography}{let1}



\bibitem{Goedgebuer} 
  {\sc J. P. Goedgebuer, L. Larger, H. Porte}, Optical cryptosystem
  based on synchronization of hyperchaos generated by a delayed
  feedback laser diode, {\it Phys. Rev. Lett.}, {\bf 80} (10),
  2249-2252 (1998).
  
\bibitem{Ikeda} 
  {\sc K. Ikeda}, Multiple-valued stationary state and its instability
  of the transmitted light by a ring cavity system, {\it Optics
    Communications}, {\bf 30} (2), 257-261 (1979).
  
\bibitem{Farmer} {\sc J. D. Farmer}, Chaotic attractors of
  infinite-dimentional dynamical system, {\it Physica D}, {\bf 4},
  366-393 (1982).



\bibitem{Noe} 
  {\sc R. Noe, U. Rückert, Y.T. Achiam, H. Porte}, European "synQPSK"
  Project: Toward Synchronous Optical Quadrature Phase Shift Keying
  with DFB Lasers, {\it OSA, Amplifiers and Their Applications/COTA},
  pp. CThC4 (2006).


\bibitem{Larger} {\sc L. Larger, J. P. Geoddgebuer, and V. Udaltsov},
  Ikeda-based nonlenear delayed dynamics for application to secure
  optical transmission systems using chaos, {\it C. R. Physique}, {\bf
    5}, 669-681 (2004).

\bibitem{Pecora} {\sc L. M. Pecora, T. L. Carroll}, Synchronization in
  chaotic systems, {\it Physical Review Letters}, {\bf 64} (8),
  821-824 (1990).



\end{thebibliography}
\end{document}